\documentclass[aps,prc,superscriptaddress,twocolumn,showpacs,nofootinbib]{revtex4-1}
\usepackage[utf8]{inputenc}
\usepackage[T1]{fontenc}
\usepackage{graphicx}   %       for graphics
\usepackage{latexsym}   %       for special symbols
\usepackage{enumerate}
\usepackage{rotating,booktabs,multirow}
\usepackage{amsmath}
\usepackage{amsfonts}
\usepackage{amssymb}
\usepackage{multirow}
\usepackage{bm}
\usepackage{subfigure}
\usepackage{xcolor}
\usepackage[colorlinks=true,linktocpage=true,linkcolor=blue,citecolor=blue,allcolors=blue]{hyperref}

\begin{document}
%============================================================================
\title{The influence of nuclear short range correlations on sub-threshold particle production in proton-nucleus collisions}

\author{Tom Reichert}
\affiliation{Institut f\"ur Theoretische Physik, Goethe-Universit\"at Frankfurt, Max-von-Laue-Strasse 1, D-60438 Frankfurt am Main, Germany}
\affiliation{Frankfurt Institute for Advanced Studies (FIAS), Ruth-Moufang-Str.1, D-60438 Frankfurt am Main, Germany}
\affiliation{Helmholtz Research Academy Hesse for FAIR (HFHF), GSI Helmholtz Center for Heavy Ion Physics, Campus Frankfurt, Max-von-Laue-Str. 12, 60438 Frankfurt, Germany}

\author{J\"org~Aichelin}
\affiliation{SUBATECH UMR 6457 (IMT Atlantique, Universit\'e de Nantes, IN2P3/CNRS), 4 Rue Alfred Kastler, F-44307 Nantes, France}
\affiliation{Frankfurt Institute for Advanced Studies (FIAS), Ruth-Moufang-Str.1, D-60438 Frankfurt am Main, Germany}

\begin{abstract}
The apparent production of (multi-)strange baryons and mesons at sub-threshold energies in proton-heavy ion collisions is a consequence of short range correlations (SRC), which have recently been observed in lepton-nucleus scattering. They may enhance the available center of mass energy of individual nucleon-nucleon collisions and allow, therefore, for sub-threshold particle production in proton-nucleus collisions. Calculations demonstrate that SRC enhance the probability for particle production at nominal sub-threshold energies up to a factor of $\times 10^3$ as compared to a simple Fermi gas model. We benchmark the idea by calculating the $\Xi^-$ multiplicity in nominal sub-threshold p+Nb collisions which compare well with the data measured by the HADES collaboration. These findings are of prime relevance for upcoming experiments at the FAIR facility, especially for the study of charmed hadrons.
\end{abstract}

\maketitle

\section{Introduction}
In elementary proton-proton collisions, the production of new particles is only possible above the production threshold $\sqrt{s}_\mathrm{thr}$ defined by the sum of the masses of the minimal composition of particles that obey all conservation laws. Surprisingly, in proton-nucleus and nucleus-nucleus reactions strange hadrons ($K^+$, $\Lambda$, $\Xi$, $\phi$, ...) are copiously produced at nominal center of mass energies (determined by the assumption that projectile and target  nucleons are at rest in their respective nuclei) well below the elementary production threshold.

The measurements of (double-)strange baryons and mesons at FOPI \cite{FOPI:2002csf,FOPI:2007btf} and HADES \cite{HADES:2009mtu,HADES:2013sfy,HADES:2017jgz,HADES:2018noy} reported an unexpected strong contribution of $\phi$ decays to the $K^-$ yield below threshold, and much more $\Xi$ baryons were measured below threshold than expected. Especially the measurement of $\Xi^-$ production in p+Nb reactions at -60 MeV below threshold by the HADES collaboration \cite{Agakishiev:2015xsa} is puzzling.

This enhancement cannot be explained by the fact that, due to Fermi motion, in proton-nucleus or nucleus-nucleus collisions more energy is available for particle production than in proton-proton collisions at the same beam energy. Sub-threshold particle production hence remains a challenging topic for theoretical studies. Whereas sub-threshold $K^+$ production measured at KaoS \cite{KaoS:1997fle,KAOS:2000ekm} could be quantitatively explained by multi-step processes \cite{Hartnack:2005tr,Hartnack:2011cn}, the origin of sub-threshold $\phi$ and $\Xi$ production is still debated \cite{Steinheimer:2015sha,Song:2020clw}. Multi-step processes have also been proposed as solutions for the $\phi$ and $\Xi$ \cite{Steinheimer:2015sha} by introducing new branching ratios for heavy nucleon resonances decaying into $N^* \rightarrow \Xi KK$. However, the decay channels of these high mass resonances are only poorly constrained \cite{ParticleDataGroup:2024cfk}. Here we propose a different, more fundamental mechanism which purely relies on a realistic description of the nucleon momentum distribution in complex nuclei by taking into account recent lepton-nucleus scattering results.

That the momentum distribution of nucleons in a nucleus deviates from a Fermi distribution is known since the late 90's \cite{Blomqvist:1998fr} but only recently triple coincidence measurements A(e,e' Np) in lepton-nucleus collisions have been performed by the CLAS collaboration \cite{CLAS:2018yvt,CLAS:2019vsb} at Jefferson Lab, which allowed to determine the reason of this deviation. The measurements indicate that a fraction of $\approx$ 20\% of the nucleons in nuclei form short range correlated $pn$ pairs (rarely also $pp$ and $nn$) having large relative momenta and small pair center of mass momenta. These nucleon-nucleon pairs are generated by the strongly attractive tensor component of the $NN$ potential at distances around 1~fm and the strong repulsive core at shorter distances. This induces the general picture that nuclei are dominated by the mean-field structure below the Fermi-momentum $k_\mathrm{F}$, and by correlated $pn$ pairs at larger momenta. Therefore the occupation number is not limited by the Fermi-momentum $k_\mathrm{F} \approx 1$ fm$^{-1}$, but exhibits a momentum distribution similar of that of deuterons, reaching momenta up to $k \approx 5$ fm$^{-1}$.

The triple coincidence measurements by the CLAS collaboration allows to determine the 4-vectors of the nucleons of the correlated pair before the collision \cite{CLAS:2018xvc} and therefore their complete kinematical determination. This allows to study the consequences of the existence of correlated $pn$ pairs in collisions of other probes with nuclei. Studies have shown that these correlated pairs can explain the EMC effect \cite{CLAS:2019vsb}, which shows that the relative momentum distribution of quarks depends on the size of the struck nucleus. In addition, $pn$ correlations play and important role for one of the goals of the upcoming FAIR facility, the study of sub-threshold particle production, in particular in proton-heavy ion collisions, to elucidate their production mechanism. 

In this letter we study the influence of these short range correlations (SRC) in complex nuclei on sub-threshold particle production in proton-heavy ion collisions. We will present our argumentation quantitatively for selected hadrons ($\Lambda$, $\phi$, $\Xi$, $\Omega$, $J/\psi$ and $\Lambda_c$), but it is clear that it can be straightforwardly generalized to more rare or exotic states. We thus provide a direct link between lepton-nucleus physics, nuclear structure and intermediate energy heavy ion collisions.

\section{Momentum distribution in nuclei}
The experimental observations of the CLAS collaboration have been studied in a mean field approach by Atti and Simula \cite{CiofidegliAtti:1995qe}. They presented a generalized convolution model for the nucleon spectral function $P(k,E)$ ($P(k,E)$ describes the joint probability to find a nucleon with momentum $k$ and removal energy $E$ in a nucleus), which is based on a factorization ansatz for the nucleus wave function and agrees well with the experimental results \cite{Song:2020clw}. In their approach (and relying on the phenomenological finding that the SRC tail behaves ``deuteron-like'') the single nucleon momentum distribution function can be separated into a Fermi-like part and a deuteron-like part as
\begin{align}
    n(k) &= n_0(k) + n_1(k),
\end{align}
where $n_0(k)$ describes the Fermi-part and $n_1(k)$ the deuteron part\footnote{The distribution $n(k)$ is normalized as $\int \mathrm{d}^3k \frac{\mathrm{d}^3n(\mathbf{k})}{\mathrm{d}^3k} = 1$.}. Ref. \cite{CiofidegliAtti:1995qe} also provides simple parametrizations of the momentum distribution $n(k)$ for the nuclei $^2$H, $^3$He, $^4$He, $^{12}$C, $^{16}$O, $^{40}$Ca, $^{56}$Fe and $^{208}$Pb. 

Here we will calculate the distribution of the center of mass energies of the nucleon-nucleon collisions $\mathrm{d}N/\mathrm{d}\sqrt{s}$ in a proton-nucleus collision (from p+$^4$He to p+$^{208}$Pb) for a given incident momentum $p_\mathrm{lab}$ of the projectile proton. The incident momentum corresponds to a nominal center of mass energy $\sqrt{s_\mathrm{NN}}$ which is calculated assuming that the target nucleon is at rest\footnote{I.e. $\sqrt{s_\mathrm{NN}} = (2m_N^2 + 2m_N(E^\mathrm{kin}_\mathrm{lab} + m_N))^{1/2}$, where $m_N$ is the nucleon mass, $E^\mathrm{kin}_\mathrm{lab}$ is the kinetic beam energy, which is related to $p_\mathrm{lab}$ via $E^\mathrm{kin}_\mathrm{lab} = (p_\mathrm{lab}^2 + m_N^2)^{1/2} - m_N$.}. We employ the parametrizations given in \cite{CiofidegliAtti:1995qe} to calculate the $\sqrt{s}$ spectrum in nuclei treated as a Fermi gas (i.e. only employing $n_0$) and including the SRC tail (i.e. employing $n_0 + n_1$).
%%%%%%%%%%%%%%%%%%%%%%%%%%%%%%%%%%%%%%%%%
\begin{figure} [t!hb]
    \centering
    \includegraphics[width=\columnwidth]{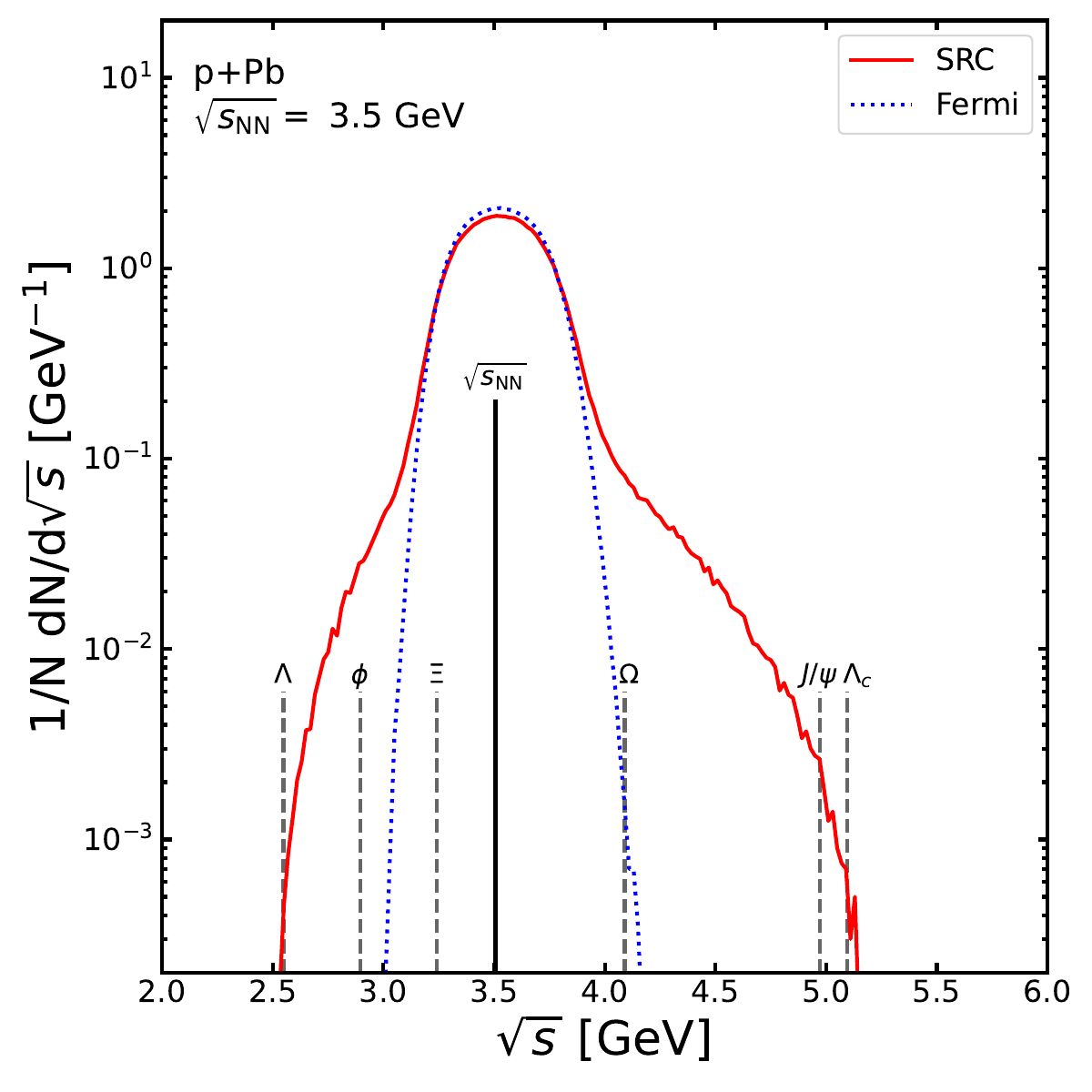}
    \caption{[Color online] The collision energy spectrum $\mathrm{d}N/\mathrm{d}\sqrt{s}$ (normalized to unity) exemplarily shown at a nominal collision energy of $\sqrt{s_\mathrm{NN}} = 3.5$ GeV in p+Pb collisions. The red solid curve shows the nucleus with SRC, the blue dotted line the nucleus without SRC. The solid vertical and dashed vertical lines show the nominal collision energy and selected hadron production thresholds.}
    \label{fig:dNdsqrts_example}
\end{figure}
%%%%%%%%%%%%%%%%%%%%%%%%%%%%%%%%%%%%%%%%%
The (normalized) collision spectrum is then given by
\begin{align} \label{eq:dNdsqrts}
    \frac{\mathrm{d}N(\sqrt{s})}{\mathrm{d}\sqrt{s}}\bigg|_{p_\mathrm{lab}} &= \int \mathrm{d}^3k \frac{\mathrm{d}^3n(\mathbf{k})}{\mathrm{d}^3k} \delta\left(\sqrt{(k^\mu + p^\mu)^2} - \sqrt{s} \right),
\end{align}
where $\mathbf{k}$ is the momentum three-vector and $k^\mu$ is the corresponding four-momentum of a nucleon with mass $m_N$ from the target nucleus, $p^\mu$ is the four-momentum of the impinging projectile nucleon defined by $p_\mathrm{lab}$ and $\mathrm{d}^3n(\mathbf{k})/\mathrm{d}^3k$ is the differential momentum distribution given by $n(k)$, where $k$ is the magnitude of $\mathbf{k}$.

From this the probability $P(\sqrt{s} \geq \sqrt{s}_\mathrm{thr})$ that an individual nucleon-nucleon collision in the pA system is above the elementary production threshold $\sqrt{s}_\mathrm{thr}$ for a specific hadron is given by
\begin{align} \label{eq:Ncoll}
    P(\sqrt{s} \geq \sqrt{s}_\mathrm{thr}) &= \int\limits_{\sqrt{s}_\mathrm{thr}}^{\infty} \mathrm{d}\sqrt{s} \frac{\mathrm{d}N}{\mathrm{d}\sqrt{s}} \\
    &\equiv \frac{N_\mathrm{coll}^{\sqrt{s} \geq \sqrt{s}_\mathrm{thr}}}{N_\mathrm{coll}^\mathrm{tot}} ,
\end{align}
where $\sqrt{s}_\mathrm{thr}$ is the threshold energy and $N_\mathrm{coll}^\mathrm{tot}$ and $N_\mathrm{coll}^{\sqrt{s} \geq \sqrt{s}_\mathrm{thr}}$ are the total number of collisions and the number of collisions above threshold, respectively. 

%%%%%%%%%%%%%%%%%%%%%%%%%%%%%%%%%%%%%%%%%
\begin{figure} [t!hb]
    \centering
    \includegraphics[width=\columnwidth]{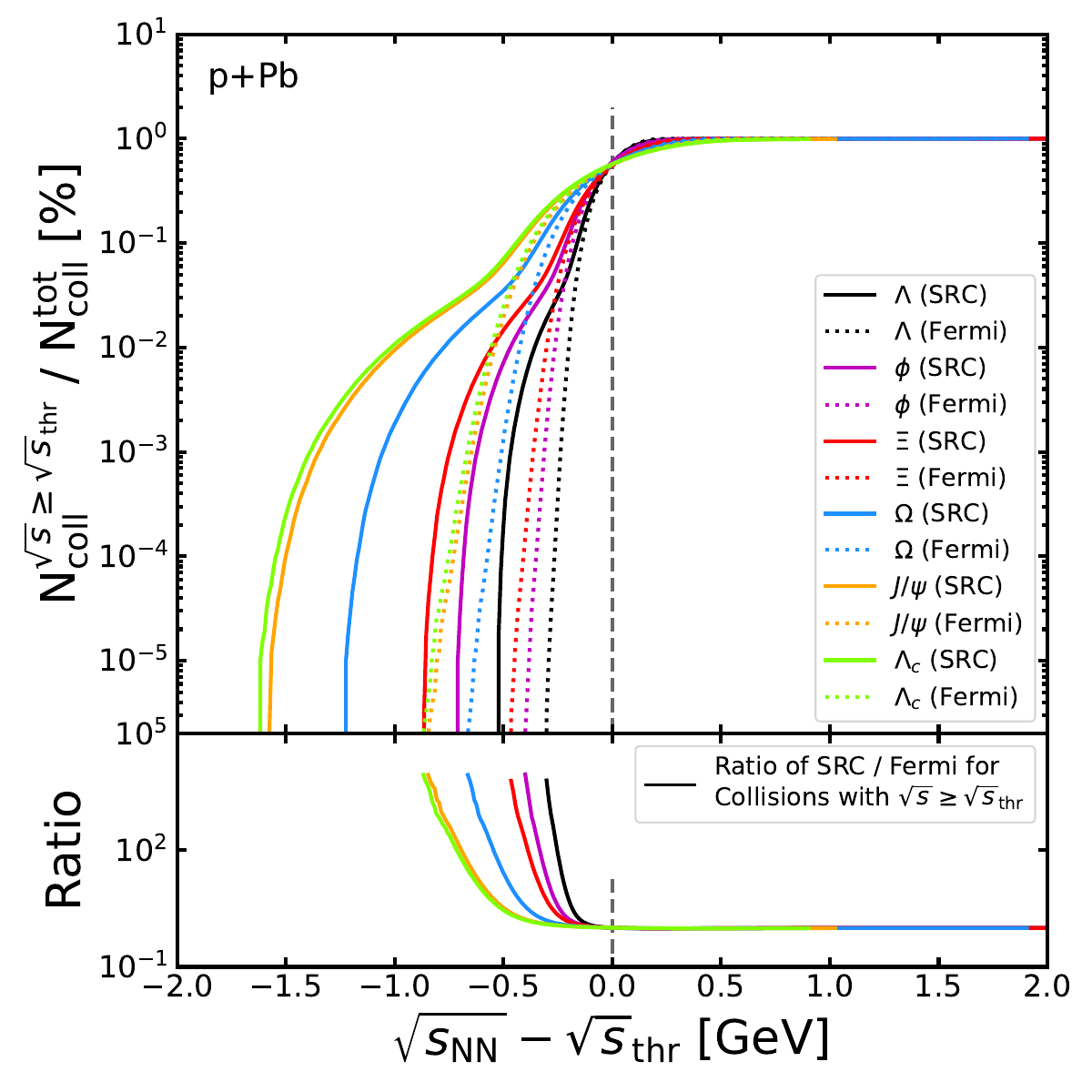}
    \caption{[Color online] The percentage of collisions above the elementary production threshold $\sqrt{s}_\mathrm{thr}$ of $\Lambda$ (black), $\phi$ (magenta), $\Xi$ (red), $\Omega$ (blue), $J/\psi$ (orange) and $\Lambda_c$ (green) in p+Pb collisions with short range correlations (solid lines) and with a simple Fermi gas (dotted lines) as a function of the reduced center-of-mass energy $\sqrt{s_\mathrm{NN}} - \sqrt{s}_\mathrm{thr}$. The lower panel shows their ratio.}
    \label{fig:shifted_coll_above_threshold}
\end{figure}
%%%%%%%%%%%%%%%%%%%%%%%%%%%%%%%%%%%%%%%%%
%%%%%%%%%%%%%%%%%%%%%%%%%%%%%%%%%%%%%%%%%
\begin{figure} [t!hb]
    \centering
    \includegraphics[width=\columnwidth]{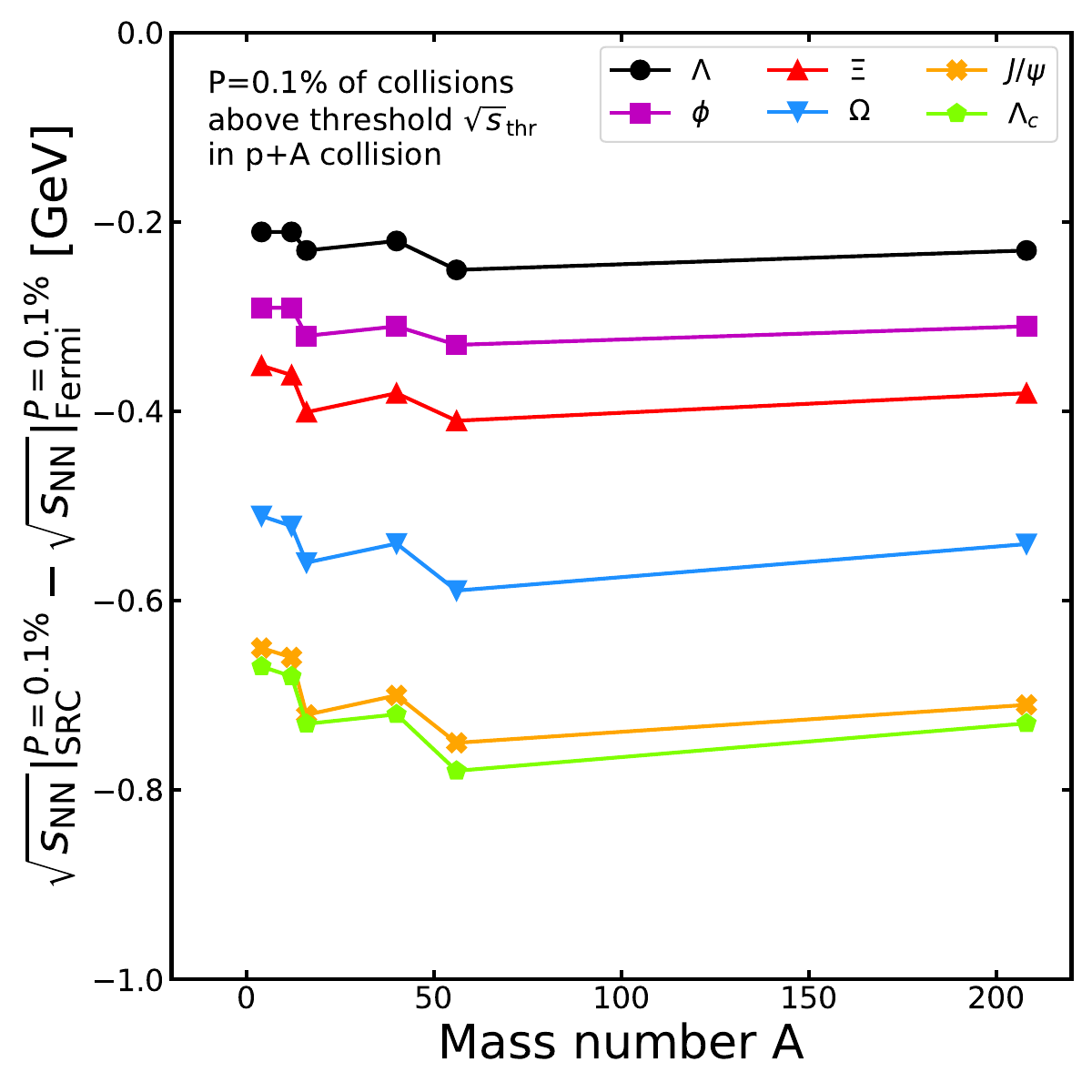}
    \caption{[Color online] Comparison of the nominal $\sqrt{s_\mathrm{NN}}$ energy, at which 0.1\% of the nucleon-nucleon collisions in a proton-nucleus collision, with (SRC) and without (Fermi) short range correlations in the target nucleus, have a $\sqrt{s} \geq \sqrt{s}_\mathrm{thr}$. In order to compare both we present their difference, i.e. $\sqrt{s_\mathrm{NN}}|^{P=0.1\%}_\mathrm{SRC} - \sqrt{s_\mathrm{NN}}|^{P=0.1\%}_\mathrm{Fermi}$. We present the results for different hadrons, respectively for different thresholds, $\Lambda$ (black), $\phi$ (magenta), $\Xi$ (red), $\Omega$ (blue), $J/\psi$ (orange) and $\Lambda_c$ (green) as a function of the mass number A of the target nucleus.}
    \label{fig:delta_sqrts_pA}
\end{figure}
%%%%%%%%%%%%%%%%%%%%%%%%%%%%%%%%%%%%%%%%%

We motivate our discussion with an exemplary calculation of the collision energy spectrum in p+Pb collisions at a nominal center of mass energy of 3.5 GeV. In Fig. \ref{fig:dNdsqrts_example} we thus exemplarily show the collision energy spectrum $\mathrm{d}N/\mathrm{d}\sqrt{s}$ (normalized to unity) at a nominal collision energy of $\sqrt{s_\mathrm{NN}} = 3.5$ GeV in p+Pb collisions. The red solid curve shows scattering on a Pb nucleus with SRC, the blue dotted line scattering on a Pb nucleus without SRC. The solid vertical and dashed vertical lines show the nominal collision energy and selected hadron production thresholds, respectively.

Here, one can clearly observe the influence that short range correlations have on the available center of mass energy for particle production. The coupling of the incoming proton to a nucleon from the deuteron-like tail leads to a much broader distribution of center of mass energies. The distribution of $\sqrt{s}$ in nuclei with SRC has a much enhanced tail at large $\sqrt{s}$ as compared to the Fermi calculation, but it also stretches further towards smaller $\sqrt{s}$ due to the random orientation of the nucleons in the nucleus\footnote{We have checked that the average $\langle \sqrt{s} \rangle$ remains mostly (within 5\%) unchanged. This is also the case for other nuclei and other nominal center of mass energies.}. One can clearly observe that the production of e.g. $\Omega$ states is strongly enhanced when one considers SRC in nuclei, as compared to the case that nucleons have only a Fermi motion.

\begin{table} [h!tb]
    \centering
    \begin{tabular}{clc}
        Particle & Channel & $\sqrt{s}_\mathrm{thr}$ [GeV] \\
        \hline 
        \hline
        $\Lambda$ & $p p \rightarrow p \Lambda K$ & 2.548 \\
        $\phi$ & $p p \rightarrow p p \phi$ & 2.896 \\
        $\Xi$ & $p p \rightarrow p \Xi K K$ & 3.241 \\
        $\Omega$ & $p p \rightarrow p \Omega K K K$ & 4.090 \\
        $J/\psi$ & $p p \rightarrow p p J/\psi$ & 4.972 \\
        $\Lambda_c$ & $p p \rightarrow p \Lambda_c D$ & 5.094 \\
        \hline
        \hline
    \end{tabular}
    \caption{The threshold production channels and energies of the $\Lambda$, $\phi$, $\Xi$, $\Omega$, $J/\psi$ and $\Lambda_c$. The masses are taken from the PDG \cite{ParticleDataGroup:2024cfk}.}
    \label{tab:thresholds}
\end{table}

%%%%%%%%%%%%%%%%%%%%%%%%%%%%%%%%%%%%%%%%%
\begin{figure} [t!hb]
    \centering
    \includegraphics[width=\columnwidth]{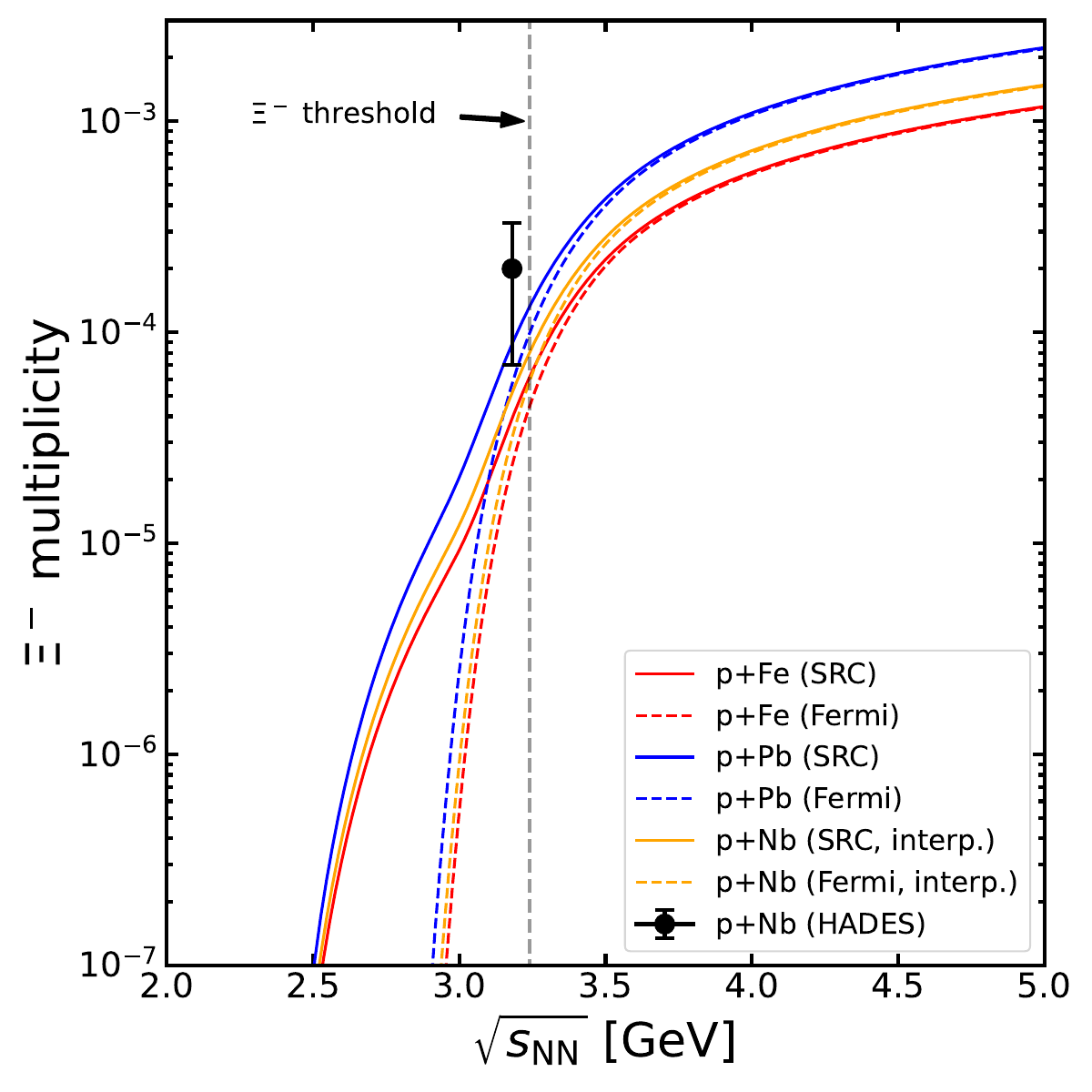}
    \caption{[Color online] The multiplicity of $\Xi^-$ baryons as a function of center of mass energy $\sqrt{s_\mathrm{NN}}$ in proton+nucleus reactions. The solid curves show the results using nuclei with SRC, while the dashed lines resemble the results using only a Fermi-distribution. The measured data point by HADES \cite{Agakishiev:2015xsa} (black circle with error bar) is measured in p+Nb reactions. We have interpolated the results from p+Fe reactions (red) and p+Pb reactions (blue) to the p+Nb case (orange). The dashed vertical line shows the threshold energy.}
    \label{fig:Xi_multiplicity}
\end{figure} 
%%%%%%%%%%%%%%%%%%%%%%%%%%%%%%%%%%%%%%%%%

\section{Sub-threshold strangeness and charm}
The unprecedented high collision rate of the SIS100 accelerator at the upcoming FAIR facility near Darmstadt, Germany yields an ideal tool for the investigation of deep sub-threshold strangeness production and sub-threshold charm production. In Tab. \ref{tab:thresholds} we list the threshold production channels and energies for the production of the $\Lambda$, $\phi$, $\Xi$, $\Omega$, $J/\psi$ and $\Lambda_c$ providing a selection of the most relevant hadrons. Their masses are taken from the PDG \cite{ParticleDataGroup:2024cfk}. In this letter we restrict ourselves to these selected hadrons, however, the calculation is straightforwardly applicable for more rare or exotic states ($\Bar{p}$, $\Xi_{cc}$, $X,Y,Z$ states, etc.) which have even larger thresholds \cite{Reichert:2025iwz}.

\subsection{Sub-threshold collisions in nuclei with SRC}
To start and benchmark the investigation we first calculate the probability (or the relative number of collisions among all collisions) that a nucleon-nucleon collision is below the production threshold (defined as the threshold in proton-proton collisions with the same kinetic energy) for the listed hadrons in p+Pb collisions as a function of the center of mass energy of the pA collision system. Later also other nuclei than Pb will be discussed.

In order to compare the threshold effect for the different hadrons, the upper panel of Fig. \ref{fig:shifted_coll_above_threshold} shows the percentage of collisions above the elementary production threshold $\sqrt{s}_\mathrm{thr}$ of $\Lambda$ (black), $\phi$ (magenta), $\Xi$ (red), $\Omega$ (blue), $J/\psi$ (orange) and $\Lambda_c$ (green) in p+Pb collisions in calculations with short range correlations (solid lines) and with a simple Fermi gas (dotted lines) as a function of the reduced center-of-mass energy $\sqrt{s_\mathrm{NN}} - \sqrt{s}_\mathrm{thr}$. The lower panel shows their ratio (i.e. the ratio of the SRC calculation to the Fermi scenario).

The calculation also shows that for a given $\sqrt{s_\mathrm{NN}}$ short range correlations enhance sub-threshold particle production as compared to the assumption that the momentum distribution of the target nucleons follows a Fermi distribution. This enhancement increases with  decreasing $\sqrt{s_\mathrm{NN}} - \sqrt{s}_\mathrm{thr}$. If one quantifies this enhancement by forming the ratio we see enhancement factors up to $10^3$ deeply below threshold, as seen in the lower panel.

The calculation also demonstrates that short range correlation allow for deeper sub-threshold production for heavier hadrons (compare $\Lambda_c$ to $\Lambda$). In case of the $\Lambda$ the relative number of collisions above the $\Lambda$ threshold drops to 0.1\% of all collisions roughly at $\sqrt{s_\mathrm{NN}} - \sqrt{s}_\mathrm{thr} \approx 0.24$~GeV for the Fermi-gas and at $\approx 0.47$~GeV including SRC, whereas for the charmed $\Lambda_c$ it drops to 0.1\% at $\approx 0.67$~GeV for the Fermi-gas and at $\approx 1.40$~GeV including SRC. 

These results also show that, despite of the limitation to a center of mass energy of $\sqrt{s} = 4.92$ GeV for Au+Au collisions at SIS100, a substantial production yield of $J/\psi$ mesons can still be expected even up to a GeV below threshold, therefore making FAIR ideally suited to study sub-threshold particle charm production in heavy ion collisions. A further major advantage of deep sub-threshold production via SRC in nuclei is that they are almost exclusively produced by pn pairs (isospin independence) and that the SRC tail in the momentum distribution is nearly identical in different nuclei. This offers the opportunity to separate the influence of SCR on sub-threshold particle production from other possible sources as e.g. multi-step processes \cite{Steinheimer:2015sha}, by measuring the target mass dependence of the sub-threshold particle production. 

Next we compare pA reactions for different nuclei. For this we calculate first the nominal center of mass energy $\sqrt{s_\mathrm{NN}}$, calculated under the assumption that the target nucleons are at rest, at which 0.1\% (we choose 0.1\% arbitrarily as the benchmark) of all collisions have a center of mass energy above the elementary threshold $\sqrt{s}_\mathrm{thr}$ for all considered hadrons, including ($\sqrt{s_\mathrm{NN}}|^\mathrm{P=0.1\%}_\mathrm{SCR}$) and excluding ($\sqrt{s_\mathrm{NN}}|^\mathrm{P=0.1\%}_\mathrm{Fermi}$) SCR. The difference between these nominal energies is plotted in Fig. \ref{fig:delta_sqrts_pA} as a function of the target mass number A for the $\Lambda$ (black), $\phi$ (magenta), $\Xi$ (red), $\Omega$ (blue), $J/\psi$ (orange) and $\Lambda_c$ (green).

The results, again, indicate that the heavier the hadron the deeper sub-threshold it can be produced. In addition one finds that the deepest sub-threshold hadron production (demanding that an equal 0.1\% of all collisions is above $\sqrt{s}_\mathrm{thr}$) is realized using Fe nuclei. Here, one gains an additional $\approx 100$ MeV in p+Fe collisions in comparison to p+He and $\approx 50$ MeV when compared to p+Pb collisions. We point out that in a heavier pA system also trivially more collisions will take place, what has been factored out here.

\subsection{$\Xi^-$ multiplicity in p+Nb reactions}
To our best knowledge, the only measured sub-threshold hadron multiplicity data in a proton-nucleus reaction are from HADES \cite{Agakishiev:2015xsa},  where the $\Xi^-$ yield in p+Nb reactions at a kinetic beam energy of $3.5$ GeV (respectively $\sqrt{s_\mathrm{NN}} = 3.18$ GeV)  and therefore -61 MeV below the elementary production threshold of the $\Xi$, has been measured. 

In our model we obtain the $\Xi$ multiplicity by multiplying the $\sqrt{s}$ spectrum with the production cross section $\sigma_{pp \rightarrow \Xi^- X}(\sqrt{s})$ and the subsequent integration over $\sqrt{s}$. The normalization of the $\sqrt{s}$ distribution has been chosen to match a Glauber calculation for the number of binary collisions.
Fig. \ref{fig:Xi_multiplicity} shows the multiplicity of $\Xi^-$ baryons as a function of center of mass energy $\sqrt{s_\mathrm{NN}}$ in proton-nucleus reactions. The solid curves show the results for nuclei with SRC, while the dashed lines resemble the results using only a Fermi-distribution. The measured data point for p+Nb by HADES \cite{Agakishiev:2015xsa}  is given by a black circle with error bar. We have interpolated the results from p+Fe reactions (red) and p+Pb reactions (blue), where parametrizations are available, to the p+Nb case (orange). The dashed vertical line shows the threshold energy.

%We first of all note that our model does not account for the dynamical evolution of the system, i.e. collecting energy via high mass resonances and in-medium effects close to threshold are not included. 
One can clearly observe that deep below threshold the estimated hadron multiplicity depends drastically on the inclusion of SRC in the nuclear momentum distribution.  The deeper below threshold the more SCR enhances sub-threshold production. Unfortunately, the pioneering measurement done by HADES \cite{Agakishiev:2015xsa} is only slightly below threshold, where the difference between the Fermi- and the SRC-calculation is on a similar magnitude as the experimental uncertainty. However, with the upcoming FAIR facility near Darmstadt it will be possible to scan $\Xi$ production much deeper below threshold in pA reactions. This will provide a direct handle on measuring SRC in strong interactions as well as a natural explanation for deep sub-threshold particle production. 

\section{Conclusion}
In this Letter we study for the first time the consequence of short range pn correlations observed in deep inelastic lepton induced eA collisions on sub-threshold particle production in proton-heavy ion collisions. Applying a parametrization, which describes the eA data well, we show that short range correlations strongly influence particle production in proton-heavy ion collisions at sub-threshold energies. The deeper sub-threshold the more the yield gets enhanced as compared to the standard Fermi gas model. This effect may well be studied at the upcoming FAIR facility and may shed light on the yet not understood sub-threshold and near threshold particle production in heavy ion collisions.

\begin{acknowledgements}
The authors thank Elena Bratkovskaya, Jan Steinheimer and Marcus Bleicher for fruitful discussion.
%T.R. thanks J.A. and the people at SUBATECH for their kind hospitality during this project.
%T.R. acknowledges support via the Procope Mobility Grant provided by the Ambassade De France En Allemagne (French Embassy in Germany) with grant number 185-DGM-E0402-03-001.
T.R. furhter acknowledges support through the Main-Campus-Doctus fellowship provided by the Stiftung Polytechnische Gesellschaft (SPTG) Frankfurt am Main and moreover thanks the Samson AG for their support.
Computational resources were provided by the Center for Scientific Computing (CSC) of the Goethe University and the ``Green Cube'' at GSI, Darmstadt. 
\end{acknowledgements}

%%%%%%%%%%%%%%%%%%%%%%%%%%%%%%%%%%%%%%%%%
\bibliography{bibliography}
%%%%%%%%%%%%%%%%%%%%%%%%%%%%%%%%%%%%%%%%%

\end{document}